\documentclass[english,aps,prb,showpacs,superscriptaddress,twocolumn,floatfix]{revtex4}
\usepackage[T1]{fontenc}
\usepackage[latin9]{inputenc}
\usepackage{graphicx}
\usepackage{amssymb}
\usepackage{babel}

\begin{document}

\title{Cavity Cooling of Mechanical Resonator in the Presence of TLS Defects}

\author{Lin Tian}

\email{ltian@ucmerced.edu}

\affiliation{5200 North Lake Road, University of California, Merced, California 95343, USA}

\begin{abstract}
Cavity cooling via quantum back-action force can extract thermal fluctuations from a mechanical resonator to reach the quantum ground state. Surface or bulk two-level system (TLS) defects in a mechanical resonator couple to the mechanical mode via deformation potential and can affect the cooling process significantly. Here, we develop a theory to study the cavity cooling of a mechanical mode in the presence of a TLS defect using the adiabatic elimination technique. Our result shows that the cooling process depends strongly on the resonance and the damping rate of the TLS. 
\end{abstract}
\maketitle

\section{Introduction}Mechanical resonators in the quantum limit can be used to explore macroscopic quantum effects and the quantum-to-classical boundaries in such systems.\cite{Blencowe} To reach the quantum ground state of the mechanical systems, many approaches have been explored, including feedback cooling, dynamic back-action cooling, and cooling via quantum bits.\cite{Kippenberg, Metzger, QubitCooling} Mechanical resonators with a wide range of frequencies and high qualify-factors have been fabricated \cite{Harris, Cleland} and the interface between mechanical modes and other quantum systems has also been studied for quantum information processing.\cite{Tian2010}

Among the various approaches to reaching the quantum ground state of a mechanical resonator, sideband cooling can be achieved by coupling the resonator to an optical or microwave cavity.\cite{Wilson-Rae2007, Marquardt2007, Genes, Clerk, Tian2009, Armour} With the cavity driven at the detuning $\Delta_{b}$, the cooling (heating) rate $\Gamma_{-}$ ($\Gamma_{+}$) can be written as
\begin{equation}
\Gamma_{\mp}=g_{0}^{2}\kappa_{0}/[\kappa_{0}^2/4+(\omega_{m}\pm\Delta_{b})^{2}]\label{eq:Gapm}
\end{equation}
where $\omega_{m}$ is the mechanical frequency, $\kappa_{0}$ is the cavity damping rate, and $g_{0}$ is the effective linear coupling between the mechanical mode and the cavity mode under the driving. The cooling rate reaches a maximum of $\Gamma_{-}=4g_{0}^{2}/\kappa_{0}$ when the cavity detuning is at the first red-sideband frequency with $-\Delta_{b}=\omega_{m}$. The cooling process corresponds to an energy up-conversion of the thermal phonons in the mechanical resonator to the cavity photons. Given the mechanical damping rate $\gamma_{m}$ and the thermal phonon number $n_{th}$, the steady state phonon number under the cavity cooling is 
\begin{equation}
n_{ss}=(\Gamma_{+}+\gamma_{m}n_{th})/[(\Gamma_{-}-\Gamma_{+})+\gamma_{m}],\label{eq:nss}
\end{equation}
which is ultimately limited by the heating rate $\Gamma_{+}$. In recent experiments, the resolved-sideband regime with $\kappa_{0}\ll\omega_{m}$ has been demonstrated which shows that it is promising to reach the quantum ground state in such systems.\cite{Kippenberg2009, Wang, Schliesser, Rocheleau, Teufel}

Molecular adsorbates and crystal defects exist on the surface or in the bulk of mechanical resonators.\cite{Pohl, Phillips, Anderson} These structures can be modeled as two-level systems (TLS) and their acoustic and thermodynamic properties have been extensively studied in amorphous solids.\cite{Seoanez, Neeley} Recently, it was experimentally demonstrated that both the mechanical resonance and mechanical damping can be affected strongly by the TLS defects.\cite{Kuhn, Kippenberg2010} The vibration of a mechanical resonator modulates the asymmetric energy of a TLS via the deformation potential, and hence generates a coupling between the TLS and the mechanical mode. In a recent work,\cite{Remus} this coupling and its effect on the decoherence of the mechanical resonator have been thoroughly studied. In this work, we study the cavity cooling of a mechanical resonator in the presence of a TLS using the adiabatic elimination technique that is widely exploited in quantum optics. The energy spectrum of the coupled resonator-TLS system contains polariton states that are quite different from the harmonic oscillator spectrum of a bare mechanical mode.\cite{Blais} The cooling process, strongly depending on the energy spectrum, can hence be affected significantly. As we will show, the ``simple'' approach of adding the TLS and its coupling directly to the cooling equation of the mechanical resonator does not describe the cooling accurately.  We derive the master equation for the resonator-TLS system and study how the steady state will be affected by the properties of the TLS. This theory can be extended to study the cooling of a mechanical resonator coupled with more complicated structures such as multiple TLS's. The theory can also be extended to include the dynamics of the TLS in the cooling process. Our results can be used to analyze sideband cooling schemes for mechanical resonators when taking into account the defects. This paper is organized as the following.  In Sec.~\ref{sec:system}, we present the Hamiltonian of the coupled system including the mechanical mode, the TLS, the cavity used in the cooling process, and the bath modes of the thermal reservoirs for each of the sub-systems. Then, in Sec.~\ref{sec:cooling}, we derive the cooling master equation for the coupled resonator-TLS system using the adiabatic elimination technique. The results from this master equation and the consequence of the presence of the TLS on the cooling will be discussed in detail in Sec.~\ref{sec:results}. Finally, discussions and conclusions will be presented in Sec.~\ref{sec:conclusion}.

\section{The Coupled System\label{sec:system}}
\begin{figure}
\includegraphics[width=8.5cm,clip]{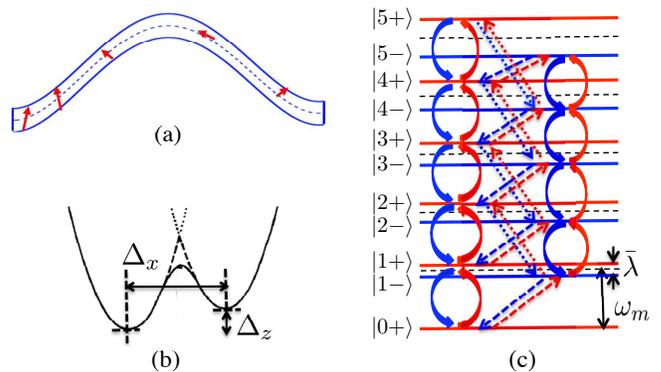}
\caption{\label{fig1}(Color online) (a) Mechanical resonator couples with TLS defects. (b) Double-well potential model for the TLS. (c) Energy spectrum of the coupled system. The eigenstates are the polariton doublets labeled as $|n\pm\rangle$.  The solid (dashed) arrows indicate transitions between states of identical (opposite) polarizations.}
\end{figure}

Our system consists of the mechanical resonator, the TLS defect, and a cavity that couples with the resonator via, e.g. optomechanical force.\cite{Kippenberg}  As is shown in Fig.~\ref{fig1}, the TLS couples with the resonator with the Hamiltonian 
\begin{equation}
H_{\tau}=\hbar\omega_{m}a^{\dag}a+\hbar[\Delta_{z}/2+\lambda(a+a^{\dag})]\sigma_{z}+(\hbar\Delta_{x}/2)\sigma_{x}\label{eq:Htau}
\end{equation}
where $a$ ($a^{\dag}$) is the annihilation (creation) operator for the mechanical mode, $\sigma_{x,z}$ are the Pauli operators, $\Delta_{z}$ is the asymmetric energy of the TLS, and $\Delta_{x}$ is the tunneling matrix element between the two sites of the TLS. The mechanical displacement generates a strain tensor in the location of the defect, and hence generates a coupling between the resonator and the TLS. The coupling amplitude $\lambda$ is proportional to the deformation potential and the second order derivative of the mechanical displacement and has been studied in detail in previous work.\cite{Seoanez, Remus} For convenience of discussion, we rewrite the Hamiltonian as
 \begin{equation}
H_{\tau}=\hbar\omega_{m}a^{\dag}a+(\hbar\omega_{z}/2)\bar{\sigma}_{z}+\hbar\bar{\lambda}(a\bar{\sigma}_{+}+a^{\dag}\bar{\sigma}_{-})\label{eq:Htau2}
\end{equation}
in terms of the rotated Pauli matrices $\bar{\sigma}_{x,z}$ with
\begin{equation}
\bar{\sigma}_{z}=(\Delta_{z}/\omega_{z})\sigma_{z}+(\Delta_{x}/\omega_{z})\sigma_{x},
\end{equation}
where $\omega_{z}=\sqrt{\Delta_{z}^{2}+\Delta_{x}^{2}}$ is the frequency of the TLS and $\bar{\lambda}=\lambda (\Delta_{x}/\omega_{z})$ is the coupling constant projected in the rotated basis. Here, we have applied the rotating wave approximation to omit the term $\lambda(\Delta_{z}/\omega_{z})(a+a^{\dag})\bar{\sigma}_{z}$ and the counter rotating terms which only have higher order effects on the cooling process. The Hamiltonian in Eq.~(\ref{eq:Htau2}) has the form of the Jaynes-Cummings Model \cite{Blais, Raimond} that has an energy spectrum distinctly different from the phonon spectrum of a bare mechanical mode as is illustrated in Fig.~\ref{fig1}. When adding the cavity mode into this system, we consider a strong red-detuned driving applied to the cavity which generates an effective linear coupling between the mechanical mode and the cavity mode.\cite{Wilson-Rae2007, Tian2009} The total Hamiltonian can be written as
\begin{equation}
H_{t}=H_{\tau}+(-\hbar\Delta_{b})b^{\dagger}b+\hbar g_{0}(a+a^{\dagger})(b+b^{\dagger})\label{eq:Ht}
\end{equation}
where $\Delta_b$ is the cavity detuning and $b$ ($b^{\dag}$) is the annihilation (creation) operator of the cavity mode. 

In addition to the quantum components,  each sub-system is subject to the environmental noise from a thermal bath which plays an essential role in the cooling process. For example, the cavity couples with a continuous spectrum of bath modes $\{b_{k}\}$ with the coupling Hamiltonian
\begin{equation}
H_{cn}=\sum_k (c_k b^\dag b_{k}+c_k^{\star}b_{k}^\dag b)\label{eq:Hcn}
\end{equation}
where $c_k$ is the coupling coefficient. The coupling with the bath modes induces cavity damping with the damping rate $\kappa_{0}$. The mechanical resonator and the TLS also couple with bath modes in similar forms as that of $H_{cn}$.   At the moment, no definite theory or experimental results are available to accurately describe the TLS decoherence. For simplicity, we will model the environment of the TLS as a thermal reservoir with the damping rate $\gamma_\tau$.  Similarly, the intrinsic damping rate of the mechanical mode is assumed to be $\gamma_m$.

\section{Cavity Cooling of the Coupled System\label{sec:cooling}}
In this section, we study the cooling of the resonator-TLS system via a cavity mode driven by red-detuned source. When the coupling between the resonator and the TLS is strong, the effect of the coupling on the eigenenergy spectrum of the coupled system can strongly affect the cooling process. We will study the cooling process using a master equation approach and derive the cooling equation by the adiabatic elimination technique.

\subsection{Eigenbasis}
The nonzero coupling between the mechanical resonator and the TLS modifies the eigenenergy spectrum of  the resonator-TLS system, as is shown in Fig.~\ref{fig1}c. Let $|\uparrow,\downarrow\rangle$ be the eigenstates of the $\bar{\sigma}_{z}$ operator of the TLS and $|n\rangle$ be the Fock states of the mechanical mode. The eigenstates of the coupled system include the ground state $|0\downarrow\rangle$ and the polariton doublets \cite{Blais} 
\begin{equation}
|n\alpha\rangle=c_{\alpha}^{n}|n\downarrow\rangle+s_{\alpha}^{n}|(n-1)\uparrow\rangle \label{eq:nalpha}
\end{equation}
with $n\ge1$ and $\alpha=\pm$. The coefficients of the eigenstates are 
\begin{eqnarray}
c_{+}^{n}=-s_{-}^{n}=\cos(\delta_{n}/2) \nonumber \\
s_{+}^{n}=c_{-}^{n}=\sin(\delta_{n}/2) \label{coeff}
\end{eqnarray}
where $\cos(\delta_{n}/2)=\sqrt{(\omega_{tn}+\delta\omega)/2\omega_{tn}}$, $\delta\omega=\omega_{m}-\omega_{z}$ is the off-resonance between the resonator and the TLS, and $\omega_{tn}=\sqrt{\delta\omega^{2}+ 4\bar{\lambda}^{2}n} $. The eigenenergies of the polariton states are 
\begin{equation}
\omega_{n\alpha}=n\omega_m +(\alpha \omega_{tn}- \delta\omega)/2
\end{equation}
for the $n$-th doublet. 

In the following, we write the quantum operators of each sub-system in terms of the eigenbasis of the coupled resonator-TLS system. For example, the mechanical annihilation operator $a$, which appears in the coupling between the cavity and the mechanical resonator in Eq.~(\ref{eq:Ht}), can be written in the eigenbasis as $a=\sum A_{\beta\alpha}^{(n)}\hat{O}_{n}^{\alpha\beta}$ with the operator
\begin{equation}
\hat{O}_{n}^{\alpha\beta}=|(n-1)\beta\rangle\langle n\alpha|
\end{equation}
defined with the polariton states, and the matrix elements 
\begin{equation}
A_{\beta\alpha}^{(n)}=\sqrt{n}c_{\beta}^{n-1}c_{\alpha}^{n}+\sqrt{n-1}s_{\beta}^{n-1}s_{\alpha}^{n}.\label{eq:Aba}
\end{equation}
Similarly, the spin operators for the TLS can be written as $\bar{\sigma}_{\pm}=\sum \sigma_{\beta\alpha}^{(n)}\hat{O}_{n}^{\alpha\beta}$ with $\sigma_{\beta\alpha}^{(n)} = c_{\beta}^{n-1} s_{\alpha}^{n}$. 

The total master equation for this system including the cavity mode can be written as \cite{QO} 
\begin{equation}
\frac{d\rho}{dt} = -\frac{i}{\hbar}[H_{t},\,\rho]+\frac{\kappa_{0}}{2}{\cal L}(b)\rho +
 \sum_{n,\alpha,\beta} \frac{\Gamma_0^{n\alpha\beta}}{2}{\cal L}_{0}^{n\alpha\beta}\rho\label{eq:rho}
\end{equation}
where ${\cal L}(o)\rho$ is defined as the Lindblad form for the operator $o$ with
\begin{equation}
{\cal L}(o)\rho=2 o\rho o^\dag - \rho o^\dag o - o^\dag o\rho,
\end{equation}
and the term
\begin{equation}
{\cal L}_{0}^{n\alpha\beta}=(n_{th}^{n\alpha\beta}+1){\cal L}(\hat{O}_{n}^{\alpha\beta})+n_{th}^{n\alpha\beta}{\cal L}(\hat{O}_{n}^{\alpha\beta\dag})
\end{equation}
describes the damping between the states $|(n-1)\beta\rangle$ and $|n\alpha\rangle$ generated by the intrinsic noise reservoirs for the mechanical resonator and for the TLS. Given a flat spectrum for the intrinsic noise reservoirs with the mechanical damping rate $\gamma_{m}$ and the TLS damping rate $\gamma_{\tau}$ respectively, we can derive 
\begin{equation}
\Gamma_0^{n\alpha\beta}=|A_{\beta\alpha}^{(n)}|^{2} \gamma_m+|\sigma_{\beta\alpha}^{(n)}|^{2}\gamma_{\tau}
\end{equation}
where $n_{th}^{n\alpha\beta}=(\exp{(\hbar\omega_{n\alpha\beta}/k_{B}T)}-1)^{-1}$ is the thermal occupation number for the energy separation 
\begin{equation}
\omega_{n\alpha\beta}=\omega_{n\alpha}-\omega_{(n-1)\beta}\label{omn}
\end{equation}
between the states $|(n-1)\beta\rangle$ and $|n\alpha\rangle$.

\subsection{Adiabatic Elimination and Master Equation}
Let $\rho_{\tau}=\textrm{Tr}_{c}(\rho)$ be the reduced density matrix for the coupled resonator-TLS system by tracing over the cavity mode. The cooling master equation for the reduced density matrix can be derived by applying the adiabatic elimination technique \cite{Cirac} to Eq.~(\ref{eq:rho}) in the limit of $\kappa_{0}\gg g_{0}, \gamma_{m}, \gamma_{\tau}$. Under strong cavity damping, only density matrix components with low photon numbers: $\rho_{\tau}^{(mn)}=\langle m_{c}|\rho|n_{c}\rangle$ for $m_{c}, n_{c}=0, 1$, need to be considered. After integrating over the time variable $t$ for the duration $t>1/\kappa_{0}$,  it can be shown that  $\rho_{\tau}^{(01)}$,  $\rho_{\tau}^{(10)}$, and $\rho_{\tau}^{(11)}$ adiabatically follow the time evolution of $\rho_{\tau}^{(00)}$.  The terms $\rho_{\tau}^{(01)}$ and $\rho_{\tau}^{(10)}$ are of the first order of the small ratio $g_{0}/\kappa_{0}$ and the term $\rho_{\tau}^{(11)}$ is of the second order of $g_{0}/\kappa_{0}$.  Hence, keeping all the terms to the second order of  $g_{0}/\kappa_{0}$, we can use the time evolution of $\rho_{\tau}^{(00)}$ to approximate the cooling master equation for $\rho_{\tau}$. We derive
\begin{eqnarray}
\frac{d\rho_{\tau}}{dt} &=& -i[\widetilde{H}_\tau,\rho_{\tau}] /\hbar
+ \sum_{n,\alpha,\beta} \frac{\Gamma_0^{n\alpha\beta}}{2}{\cal L}_{0}^{n\alpha\beta}\rho_{\tau}   \label{eq:rhot}  \\ 
&+& \sum_{n,\alpha,\beta} |A_{\beta\alpha}^{(n)}|^{2}[\frac{\Gamma_{-,\alpha\beta}^{n}}{2} {\cal L}(\hat{O}_{n}^{\alpha\beta}) + \frac{\Gamma_{+,\alpha\beta}^{n}}{2}{\cal L}(\hat{O}_{n}^{\alpha\beta\dag})]\rho_{\tau}  \nonumber
\end{eqnarray}
as the master equation for the coupled system. The first term in the equation includes the polariton Hamiltonian $\widetilde{H}_\tau=\sum_{n,\alpha }\hbar \widetilde{\omega}_{n\alpha} |n\alpha\rangle\langle n\alpha|$ with the modified polariton frequencies $\widetilde{\omega} _{n\alpha}$.  The modified frequencies include small corrections of the second order of $g_{0}/\kappa_0$. The second term describes the decoherence due to the intrinsic noise reservoirs for the mechanical resonator and the TLS, as is given in Eq.~(\ref{eq:rho}). The last term in the above equation describes the cavity cooling in the eigenbasis with the cooling (heating) rates 
\begin{equation}
\Gamma_{\mp,\alpha\beta}^{n}=g_{0}^{2}\kappa_{0}/[\kappa_{0}^2/4+(\omega_{n\alpha\beta}\pm\Delta_{b})^{2}],\label{eq:Ga12}
\end{equation}
where $\omega_{n\alpha\beta}$ is defined in Eq.~(\ref{omn}).  

The main differences between the above master equation and the standard cooling equation \cite{Wilson-Rae2007, Marquardt2007} are: 1. the cooling rates in Eq.~(\ref{eq:Ga12}) depend on the energy differences $\omega_{n\alpha\beta}$ which are state-dependent and are modified by the finite coupling constant $\bar{\lambda}$, compared with Eq.~(\ref{eq:Gapm}); 2. the matrix elements $A_{\beta\alpha}^{(n)}$ which is the projection of the annihilation operator $a$ in the polariton basis can be \emph{quite} different from the factor $\sqrt{n}$ in the cooling equation for a bare mechanical mode. As we will see below, the matrix elements $A_{\beta\alpha}^{(n)}$ can play an essential role in the cooling process.

For $\bar{\lambda} =0$, it can be shown that Eq.~(\ref{eq:rhot}) recovers the form of the cooling equation for a bare resonator.\cite{Wilson-Rae2007, Marquardt2007} For a finite $\bar{\lambda}$,  the cooling process can be significantly altered. In the dispersive regime with $|\omega_m-\omega_z|\gg\bar{\lambda}$, $A_{\alpha\alpha}^{(n)}\approx\sqrt{n}$ for the transitions between states with identical polarization, but $A_{+-}^{(n)} = (\bar{\lambda}/\delta\omega)$ and $A_{-+}^{(n)}=O[(\bar{\lambda}/\delta\omega)^3]\sim0$ for the transitions between states with opposite polarization. The mechanical cooling therefore includes two separate cooling ladders each involving states with identical polarization, plus an additional cooling for the TLS with a much smaller cooling rate $\sim\Gamma_{-}(\bar{\lambda}/\delta\omega)^{2}$, as is shown in Fig.~\ref{fig1}c. This is further confirmed by an analytical study of the cooling process in the dispersive regime where the cooling can be studied by applying the unitary transformation \cite{Blais}
\begin{equation}
U=\exp{[-(a\sigma_{+}-a^{\dag}\sigma_{-})\bar{\lambda}/\delta\omega]}
\end{equation}
to Eq.~(\ref{eq:rho}). After the transformation, the TLS becomes decoupled from the resonator but obtains an extra coupling with the cavity mode: 
\begin{equation}
H_{\tau,c}=g_{0}(\bar{\lambda}/\delta\omega)(\sigma_{+}a+\sigma_{-} a^{\dag})
\end{equation}
which is to the first order of the factor $\bar{\lambda}/\delta\omega$. This coupling generates cooling (polarization) of the TLS with a cooling rate $\sim\Gamma_{-}(\bar{\lambda}/\delta\omega)^{2}$. The cooling of the mechanical mode also recovers the results in Eqs.~(\ref{eq:Gapm}) and (\ref{eq:nss}). Detailed comparison shows that this analytical result agrees with the numerical solution for the steady state of Eq.~(\ref{eq:rhot}). In contrast, in the near-resonance regime with $|\omega_m-\omega_z|\sim0$,  $A_{\alpha\pm\alpha}^{(n)} = (\sqrt{n}\pm\sqrt{n-1})/2$ for $n\ge2$ and $A_{1\pm1}^{(1)}=1/\sqrt{2}$. For large $n$, $A_{\alpha-\alpha}^{(n)}\rightarrow1/4\sqrt{n}$, indicating the vanishing of the transitions between states with opposite polarization. While for small $n$, the transition matrix elements between states with identical and opposite polarizations are comparable, indicating a strong mixing of all the low-lying states in the eigenbasis.

\section{Results\label{sec:results}}
To illustrate the effect of the TLS on the cooling process, we numerically solve the steady state of Eq.~(\ref{eq:rhot}). The dependences of the steady state phonon number of the mechanical mode and the TLS polarization on the TLS frequency, the TLS damping rate, and the cavity detuning are studied. 

The parameters we choose are comparable with what have been achieved in recent experiments using microwave or optical cavities:\cite{Kippenberg2009, Blais} $\omega_{m}=200\, \textrm{MHz}$, mechanical damping rate $\gamma_{m}/\omega_{m}=10^{-6}$ ($Q_{m} = 10^{6}$), frequency of the TLS in the range of $\omega_{z}/\omega_{m}\in(0.5,\,1.5)$, TLS damping rate in the range of $\gamma_{\tau}/\omega_{m}\in 5\times (10^{-8},\,10^{-4})$, $\bar{\lambda}/\omega_{m}=0.05$,  $g_{0}/\omega_{m}=0.05$, $\kappa_{0}/ \omega_{m} = 0.15$, and cavity detuning in the range of $-\Delta_{b}/\omega_{m}\in (0.5,\,1.5)$. The coupling constants $g_{0}$ and $\bar{\lambda}$ are chosen by estimating the geometry and material properties of the resonator.\cite{Remus}  Given a bath temperature of $\sim 100\,\textrm{mK}$, the initial thermal phonon number can be $\approx 10$. 

\subsection{Dependence on TLS Frequency and Damping}
\begin{figure}
\includegraphics[width=8.5cm,clip]{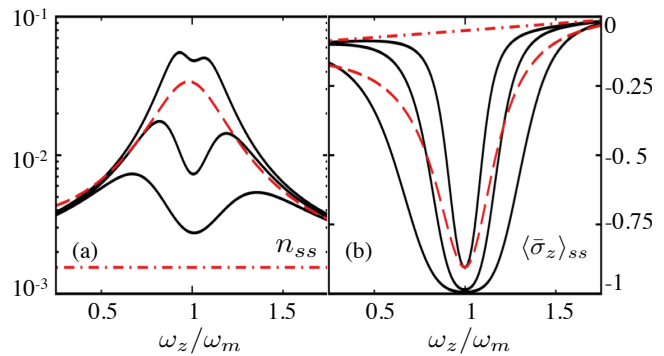}
\caption{\label{fig:2}(a) $n_{ss}$ and (b) $\langle\bar{\sigma}_{z}\rangle_{ss}$ versus the ratio $\omega_{z}/\omega_{m}$ for $\gamma_{\tau}/\omega_{m}=2.5\times (10^{-4},10^{-5},10^{-6})$ from top to bottom. Dash-dotted curves: results for zero coupling $\bar{\lambda}=0$; dashed curves: results from the ``simple'' approach (see text in Sec.~\ref{sec:conclusion}) for $\gamma_{\tau}=2.5\,\omega_{m}\times 10^{-4}$.}
\end{figure}
We first study the steady state phonon number $n_{ss}$ and spin polarization $\langle\bar{\sigma}_{z}\rangle_{ss}$ as a function of the TLS frequency $\omega_{z}$. The cavity detuning is set to be at the first red sideband frequency with $-\Delta_{b}=\omega_{m}$. In Fig.~\ref{fig:2}a, the steady state phonon number $n_{ss}$ is plotted. It can be seen that the mechanical cooling can be significantly degraded as the TLS frequency approaches the mechanical frequency. With a moderate TLS damping rate of $\gamma_{\tau}/\omega_{m} =2.5\times10^{-4}$ and a cryogenic temperature of $k_{B}T=10\,\hbar\omega_{m}$, $n_{ss}$ can be raised by nearly $50$ times due to the presence of the TLS when $\omega_z=\omega_m$; while $n_{ss}$ is only slightly increased when $\omega_z=0.5\, \omega_m$. When the two frequencies are in resonance, the thermal noise from the reservoir of the TLS can be effectively transferred to the mechanical resonator and causes severe heating of the mechanical mode. When there is a finite off-resonance between the two frequencies, the off-resonance generates an energy barrier that prevents the transfer (exchange) of energy between the two sub-systems. Hence, in the dispersive regime, the mechanical cooling is only slightly degraded. 

Meanwhile, the cavity cooling process extracts and damps the thermal noise in the TLS via its coupling with the mechanical mode. In Fig.~\ref{fig:2}b, the TLS polarization $\langle\bar{\sigma}_{z}\rangle_{ss}$ is plotted. The TLS is maximally polarized to approach the state $|\downarrow\rangle$ when $\omega_z\sim\omega_m$. With the above parameters, we have $\langle\bar{\sigma}_{z}\rangle_{ss}\approx-0.9$ at $\gamma_{\tau}=2.5\,\omega_{m}\times 10^{-4}$. A finite off-resonance prevents effective extraction of the thermal noise in the TLS in the dispersive regime. In Fig.~\ref{fig:2}a, we observe dips in the steady state phonon number $n_{ss}$ when the TLS frequency approaches the mechanical frequency $\omega_{m}$. The appearance of these dips reflects the partial reduction of the thermal noise transferred from the TLS to the mechanical resonator when the TLS is highly polarized. As a result, the heating of the mechanical mode by the presence of the TLS, although very effective as the two sub-systems are in resonance, is partially reduced.  

\begin{figure}
\includegraphics[width=8.5cm, clip]{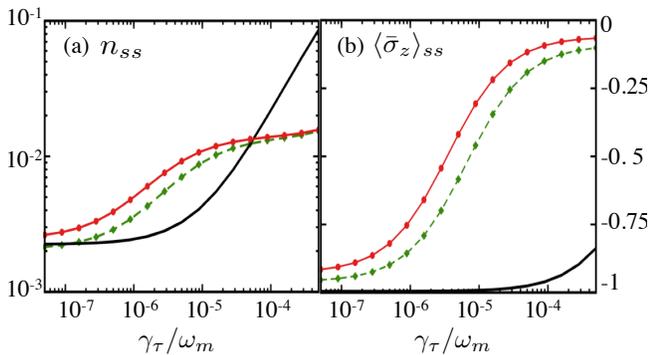}
\caption{\label{fig:2damping}(a) $n_{ss}$ and (b) $\langle\bar{\sigma}_{z}\rangle_{ss}$ versus the ratio $\gamma_{\tau}/\omega_{m}$ for $\omega_{z}/\omega_{m}=1.3, 1, 0.7$ (diamond-dashed, solid, circle-solid).}
\end{figure}
We also study the dependence of the steady state behavior on the damping of the TLS. In Fig.~\ref{fig:2damping}, the steady state phonon number $n_{ss}$ and spin polarization $\langle\bar{\sigma}_{z}\rangle_{ss}$ are plotted as a function of $\gamma_{\tau}/\omega_{m}$.  It can be seen that $n_{ss}$ increases sharply with the damping rate $\gamma_{\tau}$ when the TLS frequency is in resonance with the mechanical frequency, in contrast to the much slower increase when the two sub-systems are off-resonance.  At a damping rate $\gamma_{\tau}/\omega_{m}>10^{-6}$, the TLS is only strongly polarized when it is in resonance with the mechanical mode (the solid curve in Fig.~\ref{fig:2damping}b).

\subsection{Dependence on Cavity Detuning}
\begin{figure}
\includegraphics[width=8.5cm,clip]{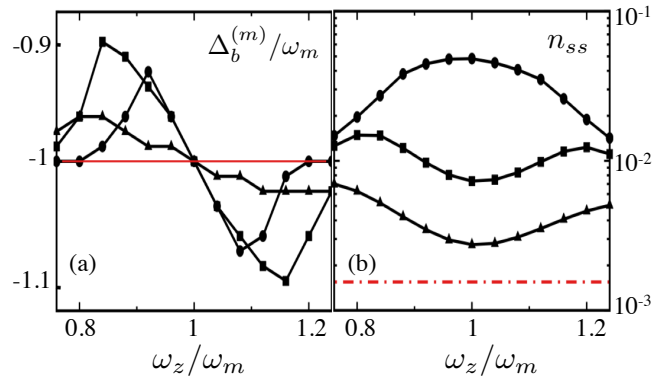}
\caption{\label{fig:3} (a) $\Delta_{b}^{(m)}/\omega_{m}$ and (b) $n_{ss}$ at cavity detuning $\Delta_{b}^{(m)}$ versus $\omega_{z}/\omega_{m}$ for $\gamma_{\tau}/ \omega_{m}=2.5\times (10^{-4},10^{-5},10^{-6})$ (circle, square, triangle) respectively. Thin straight line:  results from the ``simple'' approach; dashed curve: results for zero coupling $\bar{\lambda}=0$.}
\end{figure}

Next, we study the dependence of the steady state properties on the cavity detuning. In the cavity cooling of a bare mechanical resonator, the optimal cavity detuning $\Delta_{b}^{(m)}$ to achieve best cooling (lowest achievable $n_{ss}$) is at the red sideband frequency:  $-\Delta_{b}^{(m)}=\omega_{m}$. In our system due to the presence of the TLS, this behavior can be altered.  Using Eq.~(\ref{eq:rhot}) and studying the steady state by varying the cavity detuning, we numerically obtain the optimal detuning $\Delta_{b}^{(m)}$ as a function of the TLS frequency $\omega_{z}$, which is plotted in Fig.~\ref{fig:3}a. It can be seen that the optimal detuning is shifted away from the mechanical resonance as the TLS frequency varies, and demonstrates a non-monotonic dependence on the TLS frequency. This dependence can be explained as a combined effect due to two factors.  The first factor is the cooling of the mechanical resonator.  Without the TLS, the cooling is optimal at the detuning $-\Delta_{b}=\omega_{m}$.  The second factor is the polarization of the TLS which is optimal when the cavity detuning is at the TLS resonance with $-\Delta_{b}=\omega_{z}$. Note that the polarization of the TLS reduces the heating of the mechanical mode by the thermal noise from the TLS and hence improves the cooling of the mechanical mode.  When $\omega_{z}=\omega_{m}$, both factors reach their optimal value at the red sideband frequency, so that  we have $-\Delta_{b}^{(m)}=\omega_{m}$. However, when $\omega_{z}\ne\omega_{m}$, the optimal cooling can be reached at a cavity detuning that balances the two factors, i.e.  $-\Delta_{b}^{(m)}$ drifts to an intermediate value between $\omega_{m}$ and $\omega_{z}$.  In the deep dispersive regime, the noise from the reservoir of the TLS is largely screened from affecting the mechanical resonator. The cooling is again dominated by the first factor so that $-\Delta_{b}^{(m)}\rightarrow\omega_{m}$ returns to the mechanical resonance. In Fig.~\ref{fig:3}b,  the phonon number at the optimal detuning is plotted as a function of $\omega_{z}$, which can be compared with Fig.~\ref{fig:2}a.

\section{Discussions and conclusions\label{sec:conclusion}}
One may ask whether the above theory gives different results from the ``simple approach'' in which the TLS and its coupling are directly added into the cooling master equation for a bare mechanical resonator? To see the differences between these two approaches, we study the steady state behavior using both the  ``simple'' approach and the above theory.  In Fig.~\ref{fig:2}, the results from the ``simple'' approach are plotted as dashed curves for $\gamma_{\tau}=2.5\,\omega_{m}\times 10^{-4}$. In Fig.~\ref{fig:3}a, the results from the ``simple'' approach are plotted as the thin straight line. We find that the dependence of the phonon number $n_{ss}$ on the TLS properties can be very different in the two approaches. In particular, as seen from Fig.~\ref{fig:3}a, the optimal cavity detuning $-\Delta_{b}^{(m)}$ in the ``simple'' approach always occurs at the first red sideband frequency; while in the above theory,  $-\Delta_{b}^{(m)}$ can shift away from the red sideband frequency by as much as $10\%$ of the mechanical resonance. Note that the results from our theory agree well with the results from a brutal-force solution of the total master equation in Eq.~(\ref{eq:rho}), which further confirms the validity of this theory. 

We have discussed the effect of a single TLS on the cavity cooling and showed that the mechanical cooling can be degraded strongly when the frequency of the TLS falls within a narrow range near the mechanical resonance. Given the miniature size of the resonators, it is a reasonable scenario to assume that very few (e.g. one or two) TLS's exist in such frequency regime in the entire sample.\cite{Seoanez, Remus} Furthermore, the theory developed above can also be extended to study the cooling of a resonator coupling with multiple TLS's by deriving a master equation for the total coupled system in their eigenbasis. We can also extend the theory to include the dynamics of the TLS defects.

In conclusion, we have studied a theory for the cavity cooling of a mechanical resonator coupling with a TLS defect. The defect, subject to thermal noise, can add extra heating to the mechanical mode and affect the cooling process.  We use the adiabatic elimination technique in the eigenbasis of the coupled resonator-TLS system and derive the cooling master equation for the coupled system.  Our results showed that the cooling of the resonator can be significantly affected by the thermal noise of the TLS when the TLS frequency approaches the mechanical frequency. This theory can also be extended to describe the cooling of mechanical resonators coupling with more complicated quantum structures such as multiple TLS's. 

\section*{Acknowledgements} This work is supported by the DARPA/MTO ORCHID program through AFOSR, NSF-DMR-0956064, NSF-CCF-0916303, and COINS.


\begin{thebibliography}{10}
\bibitem{Blencowe}M. Blencowe, Phys. Rep. \textbf{395}, 159 (2004); K. C. Schwab and M. L. Roukes, Phys. Today \textbf{58}, 36 (2005); S. Bose, K. Jacobs, and P. L. Knight, Phys. Rev. A \textbf{59}, 3204 (1999).

\bibitem{Kippenberg}F. Marquardt and S. M. Girvin, Phys. \textbf{2}, 40 (2009); I. Favero and K. Karrai, Nat. Photon. \textbf{3}, 201 (2009); T. J. Kippenberg and K. J. Vahala, Science \textbf{321}, 1172 (2008).

\bibitem{Metzger}C. H. Metzger and K. Karrai, Nature \textbf{432}, 1002 (2004); O. Arcizet and \emph{et al.}, Nature \textbf{444}, 71 (2006); S. Gigan and \emph{et al.}, Nature \textbf{444}, 67 (2006); T. Corbitt and \emph{et al.}, Phys. Rev. Lett. \textbf{98}, 150802 (2007). 

\bibitem{QubitCooling}I. Wilson-Rae, P. Zoller, and A. Imamoglu, Phys. Rev. Lett. \textbf{92}, 075507 (2004); I. Martin, A. Shnirman, L. Tian, and P. Zoller, Phys. Rev. B \textbf{69}, 125339 (2004).

\bibitem{Harris}J. D. Thompson and \emph{et al.}, Nature \textbf{452}, 72 (2008).

\bibitem{Cleland}A. D. O\textquoteright{}Connell and \emph{et al.}, Nature \textbf{464}, 697 (2010).

\bibitem{Tian2010}D. E. Chang and \emph{et al.}, Proc. Natl. Acad. Sci. \textbf{107}, 1005 (2010);  L. Tian and H. L. Wang, Phys. Rev. A \textbf{82}, 053806 (2010).

\bibitem{Wilson-Rae2007}I. Wilson-Rae, N. Nooshi, W. Zwerger, and T. J. Kippenberg, Phys. Rev. Lett. \textbf{99}, 093901 (2007).

\bibitem{Marquardt2007}F. Marquardt, J. P. Chen, A. A. Clerk, and S. M. Girvin, Phys. Rev. Lett. \textbf{99}, 093902 (2007).

\bibitem{Genes}C. Genes and \emph{et al.}, Phys. Rev. A \textbf{77}, 033804 (2008).

\bibitem{Clerk}F. Elste, S. M. Girvin, and A. A. Clerk, Phys. Rev. Lett. \textbf{102}, 207209 (2009).

\bibitem{Tian2009}L. Tian, Phys. Rev. B \textbf{79}, 193407 (2009). 

\bibitem{Armour}D. A. Rodrigues and A. D. Armour, Phys. Rev. Lett. \textbf{104}, 053601 (2010).

\bibitem{Kippenberg2009}S. Gr\"{o}blacher and  \emph{et al.}, Nat. Phys. \textbf{5}, 485 (2009).

\bibitem{Wang}Y.-S. Park and H. Wang, Nat. Phys. \textbf{5}, 489 (2009).

\bibitem{Schliesser}A. Schliesser and  \emph{et al.}, Nat. Phys. \textbf{5}, 509 (2009).

\bibitem{Rocheleau}T. Rocheleau and \emph{et al.}, Nature \textbf{463}, 72 (2009).

\bibitem{Teufel}J. D. Teufel and \emph{et al.}, Phys. Rev. Lett. \textbf{101}, 197203 (2008).

\bibitem{Pohl}R. C. Zeller and R. O. Pohl, Phys. Rev. B \textbf{4}, 2029 (1971).

\bibitem{Phillips}W. A. Phillips, J. Low Temp. Phys. \textbf{7}, 351 (1972).

\bibitem{Anderson}P.W. Anderson, B. I. Halperin, and C. M. Varma, Philos. Mag. \textbf{25}, 1 (1972).

\bibitem{Seoanez}C. Seo\'{a}nez, F. Guinea, and A. H. Castro Neto, Phys. Rev. B \textbf{77}, 125107 (2008); and references there in.

\bibitem{Neeley}M. Neeley and \emph{et al.}, Nat. Phys. \textbf{4}, 523 (2008); J. M. Martinis and \emph{et al.}, Phys. Rev. Lett. \textbf{95}, 210503 (2005).

\bibitem{Kuhn}T. K\"{u}hn, D. V. Anghel, Y. M. Galperin, and M. Manninen, Phys. Rev. B \textbf{76}, 165425 (2007);  A. D. Fefferman, R. O. Pohl, A. T. Zehnder, and J. M. Parpia, Phys. Rev. Lett. \textbf{100}, 195501 (2008); G. Zolfagharkhani and \emph{et al.}, Phys. Rev. B \textbf{72}, 224101 (2005).

\bibitem{Kippenberg2010}O. Arcizet and \emph{et al.}, \pra \textbf{80}, 021803(R) (2009);  R. Rivi\`{e}re and \emph{et al.}, preprint arXiv:1011.0290.

\bibitem{Remus}L. G. Remus, M. P. Blencowe, and Y. Tanaka, Phys. Rev. B \textbf{80}, 174103 (2009).

\bibitem{Blais}A. Blais and \emph{et al.}, Phys. Rev. A \textbf{69}, 062320 (2004).

\bibitem{Raimond}J.-M. Raimond, M. Brune, and S. Haroche, Rev. Mod. Phys. \textbf{73}, 565 (2001).

\bibitem{QO}D. F. Walls and G. J. Milburn, \emph{Quantum Optics}, 2nd ed., Springer-Verlag (2008).

\bibitem{Cirac}J. I. Cirac, R. Blatt, P. Zoller, and W. D. Phillips, Phys. Rev. A \textbf{46}, 2668 (1992).
\end{thebibliography}
\end{document}